\documentclass[10pt]{article}

\usepackage{amsmath,amssymb}

\usepackage{changepage}

\usepackage{textcomp,marvosym}

\usepackage{cite}

\usepackage{nameref,hyperref}

\usepackage[utf8]{inputenc}
\usepackage[T2A,T1]{fontenc}
\usepackage{tipa}
\usepackage{amsmath,amsfonts,amsthm,amssymb,bm}  
\usepackage{blindtext}
\usepackage{hyperref}
\usepackage{comment}
\usepackage{graphicx} %
\usepackage{multirow}
\usepackage{longtable}
\usepackage{subfigure}
\usepackage{subcaption}
\usepackage{caption}
\usepackage{color,soul}

\usepackage[nopatch=eqnum]{microtype}
\DisableLigatures[f]{encoding = *, family = * }

\usepackage[table]{xcolor}

\usepackage{array}

\newcolumntype{+}{!{\vrule width 2pt}}

\newlength\savedwidth

\usepackage[aboveskip=1pt,labelfont=bf,labelsep=period,justification=raggedright,singlelinecheck=off]{caption}

\makeatletter
\renewcommand{\@biblabel}[1]{\quad#1.}
\makeatother

\begin{document}
\vspace*{0.2in}

\begin{flushleft}
{\Large
\textbf\newline{Computational analysis reveals historical trajectory of East-Polynesian lunar calendars} 
}
\newline
\\
Miguel Valério\textsuperscript{1\Yinyang*},
Fabio Tamburini\textsuperscript{2\Yinyang*},
Michele Corazza\textsuperscript{3}
\\
\bigskip
\textbf{1} Departament de Prehistòria, Universitat Autònoma de Barcelona, Bellaterra, Spain
\\
\textbf{2} FICLIT, University of Bologna, Bologna, Italy
\\
\textbf{3} CIRSFID-ALMA AI, University of Bologna, Bologna, Italy
\\
\bigskip

%
%
\Yinyang These authors contributed equally to this work.

* miguel.valerio@uab.cat, fabio.tamburini@unibo.it

\end{flushleft}

\begin{abstract}
We investigate a type of lunar calendar known as  lists of the 'nights of the moon', found throughout East Polynesia, including Rapa Nui (Easter Island). Using computational methods, we analyzed the lexical and structural divergence of 49 calendric lists from all major archipelagos, each containing about 30 night names. Our results, presented as a rooted phylogenetic tree, show a clear split into two main groups: one including lists from Rapa Nui, Mangareva, and the Marquesas; the other comprising lists from New Zealand, Hawai‘i, the Cook Islands, the Austral Islands, Tahiti, and the Tuamotu. This pattern aligns with a recent alternative classification of East Polynesian languages into 'Distal' (Marquesan, Mangarevan, Rapanui) and 'Proximal' (Māori, Hawaiian, Tahitian, etc.) subgroups. Since both language and lunar calendars are symbolic systems passed down and changed within communities—and given the geographic isolation of many archipelagos—we interpret this correspondence as evidence that the early divergence of East Polynesian lunar calendars mirrors early population movements and language splits in the region. 
\end{abstract}


\section*{Introduction: The `nights of the moon' in East Polynesia}

Lunar calendars based on synodic months (complete cycles of the Moon's phases) were once widespread in Polynesia, with the year divided into 12 or 13 lunations \cite{kirch2001hawaiki}. Each lunar month followed the astronomical 29.53-day lunar cycle, which includes the phases of the new moon, the first quarter, the full moon, the last quarter, and then cycles back to the new moon. Nights, rather than days, served as the reference point. In East Polynesia, each `night of the moon' had a specific name, a tradition that extended across all islands where East-Polynesian languages are still spoken today (Table \ref{Table 1}). This \textit{naming} practice contrasts with the calendars of West Polynesia, such as those of Samoa, Tonga, and Tokelau, which consisted of lists where nights were mostly \textit{numbered} rather than named. Because the \textit{numbering} calendars of West Polynesia included a few names shared with East Polynesia (see below), their type may have been the source of the \textit{naming} system.

\begin{figure}[ht]
    \centering
    \fbox{\includegraphics[width=1.0
    \linewidth]{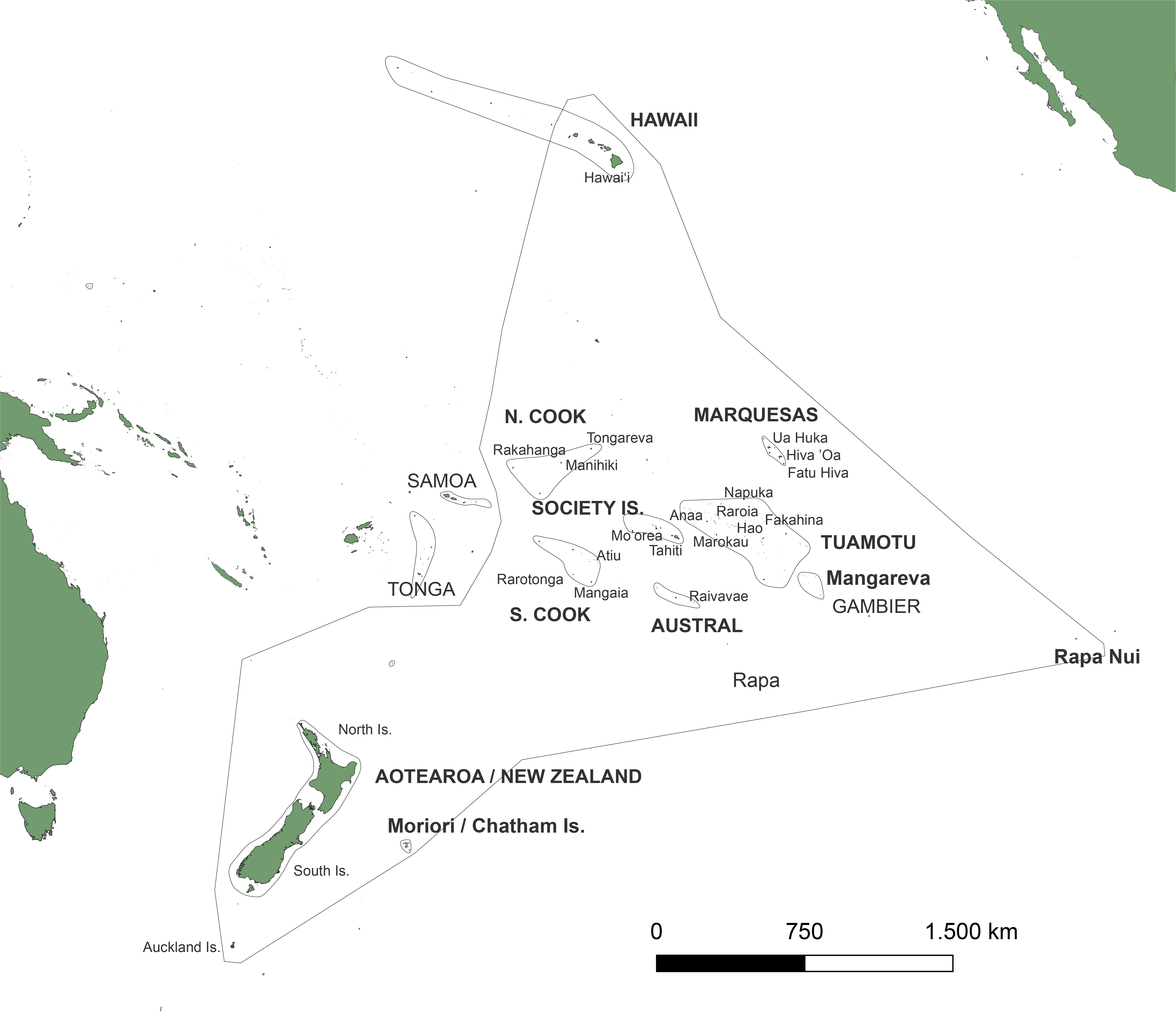}}
    \caption{The `triangle’ of East-Polynesian islands and the Samoan-Tongan area (West Polynesia). Created using the Free and Open Source QGIS.}
    \label{fig:map}
\end{figure}

East Polynesia is a large oceanic region with shared linguistic and social traits, and the last to be settled by humans. It includes the islands of Aotearoa/New Zealand, Moriori/Chatham, Society (with Tahiti), northern and southern Cook, Austral, Gambier (with Mangareva), Tuamotu, Marquesas, Rapa Nui/Easter Island and Hawai'i, among others (Fig. \ref{fig:map}). The sequence of its settlement, whose point of origin is believed to be West Polynesia (the Samoa-Tonga area), is still debated based on archaeological, anthropological, linguistic, and genetic evidence \cite{wilmshurt&al2011dating, Ioannidis2021, wilson2021EPsub}. Linguistically, East Polynesia is home to a subgroup of Polynesian tongues \cite{pawley1966, biggs1971}, spoken across islands in considerable geographical isolation. Since the typological difference between naming and numbering calendric types separates the East-Polynesian-speaking archipelagos from West Polynesia \cite{hiroa1938mgv,burrows1938western}, the prototype for the `naming' lists may have originated in a community of Proto-East Polynesian speakers.

\begin{table}[ht]
\setlength\tabcolsep{2pt}
 {\small
\hspace*{-0.65cm}\begin{tabular}[ht]{|l|l|l|l|l|}
\hline
& Rapa Nui/Easter & Hawai'i & Aotearoa/New  & Tahiti\\
& Island (RPN2) & (HAW1b) & Zealand (MAO3) & (TAH3) \\
\hline
1 & Tireo & Hilo & Whiro & Tireo \\
2 & Hiro & Hoaka & Tirea & Hiro-hiti \\
3 & Oata & Kū Kahi & Hoata & Hoata \\
4 & (O) Ari & Kū Lua & Oue & Hamiama-mua \\
5 & Kokore tahi & Kū Kolu & Okoro & Hamiama-roto \\
6 & Kokore rua & Kū Pau & Tamatea-tutahi & Hamiama-muri \\
7 & Kokore toru & 'Ole Kū Kahi & Tamatea-turua & Tamatea-mua \\
8 & Kokore ha & 'Ole Kū Lua & Tamatea-tutoru & Tamatea-roto \\
9 & Kokore rima & 'Ole Kū Kolu & Tamatea-tuwha & Tamatea-muri \\
10 & Kokore ono & 'Ole Kū Pau & Huna & Huna \\
11 & Maharu & Huna & Ari & 'Ari \\
12 & (O) Hua & Mōhalu & Maure & Maharu \\
13 & Atua & Hua & Mawharu & Hua \\
14 & (O) Hotu & Akua & Ohua & Maitu \\
15 & Maure & Hoku (Full Moon) & Atua & Hotu (Full Moon) \\
16 & Ina-ira & Mālani & Oturu (Full Moon) & Mara'i \\
17 & Rakau & Kulu & Rakau-nui & Turu \\
18 & Motohi (Full Moon) & Lā'au Kū Kahi & Rakau-matohi &
Rā'āu-mua \\
19 & Kokore tahi & Lā'au Kū Lua & Takirau &
Rā'āu-roto \\
20 & Kokore rua & Lā'au Pau & Oike & Rā'āu-muri \\
21 & Kokore toru & 'Ole Kū Kahi & Korekore-tutahi &
'Ore'ore-mua \\
22 & Kokore ha & 'Ole Kū Lua & Korekore-turua &
'Ore'ore-roto \\
23 & Kokore rima & 'Ole Kū Pau & Korekore-piri-ki-nga-Tangaroa &
'Ore'ore-muri \\
24 & Tapume & Kanaloa Kū Kahi & Tangaroa-roto &
Ta'aroa-mua \\
25 & Matua & Kanaloa Kū Lua & Tangaroa-kiokio &
Ta'aroa-roto \\
26 & (O) Rongo & Kanaloa Pau & Tangaroa-whakapau &
Ta'aroa-muri \\
27 & (O) Rongo Tane & Kāne & Otane & Tāne \\
28 & Mauri-nui & Lono & Orongonui & Ro'o-nui \\
29 & Mauri-kero & Mauli & Mauri & Ro'o-mauri \\
30 & (O) Mutu & Muku & Mutuwhenua / Omutu & Mutu \\
\hline
\end{tabular}
}
\caption{Examples of East-Polynesian lists of `nights of the moon' from Rapa Nui/Easter Island, Hawai'i, Aotearoa/New Zealand, and Tahiti.}
\label{Table 1}
\end{table}

The lunar calendar was tied to precolonial ways of life, marking suitable times for food planting and—especially—fishing. It was transmitted orally, though there is evidence that on Rapa Nui it was also inscribed on a wooden tablet in the still undeciphered Rongorongo script \cite{horley2011,valerio&al2022,valerio2024rongorongo}. After the introduction of the European solar (Gregorian) calendar and Christian festivities, this way of measuring time largely ceased to function in most places. Knowledge of it has survived because some native authorities and outsiders wrote down ``lists of the nights of the moon'' in the 19th and first half of the 20th century. In the early 1900s, a Tahitian royal person described one list as a \textit{parau} (`record') for ``fishermen, recounting the nights when the fish run, this kind and that kind, (...) and the days which are favourable for planting food-plants'' \cite{stimson1928}. Reports from various locations coincide in that some nights were ‘glossed’ orally with indications of which fishing methods (if any) were appropriate for specific moon phases, such as torchlight night fishing \cite{best1922maori,stimson1928}.

Names or parts of names for nights that repeat in all or nearly all East-Polynesian locations (cited here in their reconstructed Proto-East Polynesian forms) were likely part of the proto-calendar.  Some (e.g.\textit{*Tu(\textglotstop)u}, \textit{*Firo}, \textit{*Tamatea}, \textit{*Ta\textipa{N}aroa}, \textit{*Taane}, \textit{*Ro\textipa{N}o}, \textit{*Mauri}) match the names of deities or supernaturals \cite{craig1989dictionary,POLLEX}. Other forms seem to describe phases of the moon: \textit{*Mase\textglotstop a} `faintly perceptible', \textit{*(\textglotstop )Ari} `clearly visible', \textit{*Funa} `hidden', \textit{*Fua} `blossom', \textit{*Fotu} `appear, rise', \textit{*Matofi} `split in two', \textit{*mate} `die', \textit{Mutu} `end' \cite{POLLEX} (see S1B Appendix). Some nights formed series with the same main name (\textit{*Mase\textglotstop a \textasciitilde *Same(\textglotstop)a}, \textit{*Korekore}, \textit{*Ta\textipa{N}aroa}, etc.), followed by a word that specified the position of each homonymous night in the series. For this purpose, different systems were used, the main ones being: \textit{*tahi}, \textit{*rua}, \textit{*toru}… `one, two, three...'; \textit{*mu\textglotstop a}, \textit{*roto}, \textit{*muri} `before, inside, after'; and \textit{*tahi}, \textit{*roto}, \textit{*faka\textglotstop oti} `one, inside, final(izing)'. The numbering system for ordering series of nights may be the oldest, as it is reminiscent of the numbering lists from West Polynesia. Finally, other forms may also reflect how suitable for fishing a night was: one Tahitian account describes the \textit{*Korekore} (literally `Lack(ing)') series as nights “when the fish disappear” \cite{stimson1928}, though over time such names may have been reinterpreted in some islands. 

The archipelagos of East Polynesia used variations of the list of `nights of the moon', which, while related, show significant differences and imply divergence over time. Historical changes in the names of the nights and their sequence may be due to loss of memory and reinterpretation of the meaning of names. For instance, in the Tuamotu and the southern group of the Cook Islands,  \textit{Vari}, meaning `menstrual blood', replaced \textit{*}(\textit{\textglotstop })\textit{Ari}, possibly after the original name became semantically opaque. The duration of the lunar cycle (roughly 29.5 nights) may also have contributed to the changes. Ideally, calendric lists included 30 nights (some even contained 31 names), but in certain months the new moon would eventually appear one night earlier, which probably meant that one name had to be removed from the count \cite{hiroa1938mgv}. Such oscillations might account for the `blending' of night names seen in some calendars: e.g. \textit{Ro\textipa{N}o Taane} (Rapa Nui) combined \textit{*Ro\textipa{N}o} and \textit{*Taane}, \textit{Ro\textipa{N}o-Mauri} (several locations) combined \textit{*Ro\textipa{N}o} and \textit{*Mauri}, and \textit{Korekore-piri-Tangaroa} (New Zealand), literally `Korekore-joining-Tangaroa', occurs at the transition from the *\textit{Korekore} to the *\textit{Ta\textipa{N}aroa} series. Series were expanded or created anew and some nights were dropped. Variant names with extra descriptors, perhaps originally optional, came into use (see, e.g., \textit{*Mauri} > \textit{*Mauri-Kero} or \textit{*Mauri-mate}, both meaning roughly `Dying Mauri', when the moon was at the end of its waning phase). Another common type of divergence, more likely to be due to memory loss, is the switching of the positions of the nights: for example, we find derivatives of *(\textit{\textglotstop })\textit{Ari} both immediately before and after derivatives of \textit{*Funa} in different lists.

The similarities and differences between lists of nights of the moon from all parts of East Polynesia have long been examined comparatively to explore their relationships and common origins \cite{hiroa1932maniraka,hiroa1938mgv,metraux1940}. Early discussions mainly followed conventional and qualitative philological approaches, focusing on representative sets of lists. However, they overlooked some documented evidence and only partially addressed certain issues with the sources, such as transcription errors and interpretation of cognate relationships between night names. One early commentator \cite{hiroa1938mgv} concluded that New Zealand, the Marquesas, Mangareva, and Rapa Nui had ``a similar sequence'', despite noting some confusion, which he attributed to defective recording and disruption in oral transmission. This conclusion does not align with any current hypothesis of settlement sequence or language classification for East Polynesia. Names of nights have also been considered as evidence for the sub-grouping of East-Polynesian languages \cite{green1985,marck2000topics,walworth2014EPrev}, but strictly in terms of lexical or phonological changes, and not systematically.

To address the limitations of previous studies on the relatedness among lunar calendric lists, we applied a computational and phylogenetic method to a dataset that is, to our knowledge, more detailed and complete than any previously available in the literature. 

\section*{Materials and Methods}

\subsubsection*{Dataset}

Our dataset (S2 Dataset; sources detailed in S1 Appendix), contains 49 'lists of the nights of the Moon' from all the major archipelagos and transmitted in all East-Polynesian languages, except for the Rapan-speaking island of Rapa Iti: Aotearoa/New Zealand (n=10) and Moriori/Chatham Island (n=1), Hawai'i (n=2), Mangareva (n=1), the Marquesas (n=8), Raivavae in the Austral Islands (n=1), Rapa Nui/Easter Island (n=4), the southern (n=6) and northern Cook (n=3) groups, Tahiti and neighboring islands (n=4), and the Tuamotuan archipelago (n=9). Therefore,  it is representative of the divergence of the lists throughout East Polynesia.

\subsubsection*{Analysis of cognacy}

We define cognate names (or cognate components of names) of nights as those that (1) originate from the same etymological source and (2) occupy a similar relative position in the calendric list, regardless of whether their phonological form in a given East Polynesian language strictly follows that language's regular sound laws. This cognacy relationship is similar to the connection between the words for 'January' in different languages, such as French \textit{janvier}, German \textit{Januar}, Italian \textit{gennaio}, Portuguese \textit{janeiro}, and Spanish \textit{enero}. 

Etymologies used to determine (1) follow for the most those of POLLEX (Polynesian Lexicon Project) Online, a large-scale comparative dictionary of Polynesian languages, which uses dictionaries and other scholarly works as primary sources \cite{POLLEX}.  One notable exception involves a series of nights attested in lists from many locations, and whose names were already identified as cognates by J. F. Stimson in the 1930s: Marquesan \textit{Maheama}, Mangarevan \textit{Ma}[\textit{\textglotstop }]\textit{ema}, Tahitian \textit{Hamiāma}, Rurutu (Austral islands) \textit{\textglotstop amia}, Tongarevan \textit{Samia}, Tuamotuan \textit{Hania} or \textit{Hanīa} \cite{stimson1930hamzah}. Stimson assumed the following development: \textit{Samia} > \textit{Hamia} > \textit{Hamia-ma}(-\textit{tahi}, etc.) > \textit{Mahiama} > \textit{Maheama} / \textit{Ma}[\textit{\textglotstop }]\textit{ema}. Yet this leaves the etymology of the name unexplained, as most forms are opaque. We argue that the path was the opposite. The starting form is Proto-East Polynesian \textit{*ma-se\textglotstop }\textit{a} ‘(be) perceptible’ \cite{POLLEX}, originating the structure *\textit{mase\textglotstop }\textit{a}(\textit{-maa})-NUMERAL, where *\textit{maa} is a numeral connective particle that became lexicalized and fossilized as part of the names of nights only in some calendars. Thus, *\textit{Mase\textglotstop a}(-\textit{maa})- developed not just into Marquesan \textit{Mahea-ma} / Mangarevan \textit{Ma}[\textit{\textglotstop}]\textit{ema} (with *\textit{s} >\textit{ h}), but also into *\textit{Same}[\textit{\textglotstop}]\textit{a}(-\textit{ma}), which, following metathesis (and the irregular shift \textit{e} > \textit{i}) finally became Tongarevan \textit{Samia}, Tahitian \textit{Hamiama} (with \textit{e} >\textit{ i}), and Tuamotuan *\textit{Hamia} > \textit{Hania} (irregular \textit{m} > \textit{n}). The etymology is supported by the semantics of Hawaiian \textit{māhea} ‘hazy (moonlight)’, Rapanui \textit{ma\textglotstop eha} ‘brightness; lighten/brighten up’, and Tahitian \textit{maheahea} ‘turn pale; fade’, keeping in mind that this series of nocturnal names referred to some of the first nights after the new moon, when the latter is barely visible.

In addressing the proximity between any pair of calendric lists, we did not just compute which forms of night names, and how many, were shared (the `lexical' aspect). We also considered how much the relative position of a pair of cognate forms diverged within the sequence of ca. 30 nights (the `structural' aspect).
We analyzed cognacy among names or components of names (here referred to as `forms') of the lunar nights, rather than focusing on segments or phonemes (see next subsection). 

\subsubsection*{Notation of the data}

For compiling the dataset, our notation of the names of lunar nights from the calendric lists followed seven (1-7) principles:

 (1) \textit{The basic unit of analysis is a `form'}\textbf{,} which represents the name of a night or part thereof. This choice responds to the intricacies of the data. On the one hand, in Polynesian studies (especially in early or non-linguistic publications) the notation of morphemes varies, leading to instances where the same combination of morphemes can be spelled as one or multiple words: e.g., \textit{Rongo nui }/ \textit{Rongo-nui }/ \textit{Rongonui }`Big Rongo'. On the other hand, many night names are composite. Therefore, our unit of analysis comprises \textit{any name or part of a night name that we can examine separately for its presence or absence across calendric lists}. This unit, which we termed ‘form', comprises both full words and morphemes, as well as repeated parts of morphemes (such as `ko', which is the partial reduplication of the morpheme `kore' in `kokore'). We defined our set of forms inductively, by examining the dataset and the established relations of cognacy among night names (see the index of forms in S1B Appendix).

 (2) Forms (not character strings) were compared as whole units and notated in normalized ways—either reconstructed proto-forms or language-specific reflexes—disregarding phonological variation and linguistic change. For instance, East-Polynesian *\textit{Ro\textipa{N}o} \cite{POLLEX} appears as \textit{Ro\textipa{N}o} (Maori, Rapanui, and Tuamotuan), \textit{\textglotstop Ono}, \textit{\textglotstop Oko} (Marquesan), \textit{Lono} (Hawaiian), and \textit{Ro\textglotstop o} (Tahitian) in different lists and languages. To compute all these forms as cognates, we notated them as ‘ro\textipa{N}o’, after the proto-form. 

 (3) The `nights of the moon' were often recited using extended formulas: e.g. Tahitian \textit{\textglotstop O Ro\textglotstop o-nui te pō} ``Ro\textglotstop o-nui is the night'', Maori \textit{He Otāne} ``It is O Tāne'' (where O Tāne is literally ``[night] of Tāne''), or Rakahanga-Manihiki \textit{Ko mara\textipa{N}i} ``(The) Marangi'' (see S1A Appendix, with references). Thus, in considering only basic forms to compare the names of nights, we excluded from the analysis most grammatical words (`particles') such as markers \textit{ko} / \textit{\textglotstop o}, \textit{he}, \textit{o}, and the article \textit{te}.

 (4) Use of symbol `/': Where there are variants of a form that can be diagnostic of relatedness between lists, both `form' and `variant' were annotated for computational analysis, following the scheme FORM/VARIANT. For example, while most calendric lists have a reflex of \textit{*\textglotstop Ari}, Mangaian, Rarotongan and Tuamotuan lists have \textit{Vari}, probably due to lexical reinterpretation (as mentioned above). The presence of one or the other variant was notated as `\textglotstop ari/\textglotstop ari' vs `\textglotstop ari/vari'. The part before `/' shows that some (any) form of the night occurs, while the part after `/' part indicates the specific variant. They are distinct features, each of which may or may not be shared by any given pair of lists.

 (5) Use of symbol `\#': This was used for two purposes: (i) the comparison of lexicalized complexes of the type of FORM\#FORM against a single FORM (e.g. `mase\textglotstop a\#maa' is fully cognate with another `mase\textglotstop a\#maa' but only partially cognate with `mase\textglotstop a'); (ii) the comparison of reduplicated vs partially reduplicated vs non-reduplicated expressions of the same form; notice that reduplication is a common morphological process in Polynesian languages and it affects some night names (e.g. `kore\#kore' vs `ko\#kore' vs `kore').

 (6) Use of symbol `\_': This was used to detect full cognacy between composite night names of the type of FORM\_FORM, regardless of the fact that the forms changed syntactical position (especially in ordered series of nights). For instance, different Tuamotuan lists feature the variant names \textit{Hania fakaoti} `Hania Final' and \textit{Te fakaotiga na Hania} `The finalizing of Hania', but we treated them as cognate despite changes in word order (and exact grammatical form), so as not to generate artifacts. 

 (7) Added numbers (FORM1, FORM2, etc.) distinguish non-cognate homographic forms. To illustrate this point: some lists share the composite `ro\textipa{N}o\#nui’ while others have a standalone `ro\textipa{N}o', a subtle difference we needed to consider. Similarly, we needed to compute the greater proximity between nights called `fotu\#nui' as opposed to nights called simply `fotu'. However, we did not count as cognates `fotu\#nui' and `ro\textipa{N}o\#nui'. Even though both forms are modified by an adjective \textit{nui} `big', they are nights with quite different positions in their lists and the addition is most probably a trivial innovation that occurred independently. To put it simply, the addition of the adjective ‘big’ to \textit{*Fotu} is a separate phenomonen from its addition to \textit{*Ro\textipa{N}o}. To handle this, we added numerical labels to homographic words—both lexical and grammatical—to mark them as non-cognate forms: e.g., tuu\#nui1 vs. tamatea\_nui2 vs. fotu/fotu\#nui3, etc.; tuu\#faka1\#so\#\textglotstop ata vs faka2\#pau vs faka3\#\textglotstop oti, etc. (see S1B Appendix).

\subsubsection*{The issue of borrowability}

While the lexicon is typically the most readily borrowed component of a language—and lunar calendars might conceivably fall under this category—the degree of isolation of the East Polynesian islands makes the borrowing of individual lunar night names unlikely. However, it remains plausible that borrowing of night names and mutual influence occurred within specific archipelagos, such as among the Marquesan or the Tuamotuan islands. Furthermore, in cases where archipelagos came into contact in relatively recent times, the wholesale borrowing of calendric lists is conceivable (see below on historical contact between Tongareva and Rarotonga). Nonetheless, we argue that rather than presenting a methodological obstacle, the construction of a phylogeny of East Polynesian lunar calendars can provide a means of addressing such issues.

Accordingly, we should note at the outset that the Society, Tuamotu, southern Cook and Austral Island represent some of the least isolated archipelagos within East Polynesia. In addition, during the late 18th and early 19th centuries a newly-formed Tahitian kingdom expanded its influence in the region \cite{denoon2004cambridge}.  This facilitated increased contact across these islands. As a result, the calendric lists from Tahiti, Tuamotu, the Southern Cook, and Raivavae islands are the most likely to show evidence of horizontal transmission (borrowings), a factor we take into account when evaluating the phylogeny derived from our analysis. The case of Tongareva (Northern Cook) is also noteworthy. Although the island is geographically closest to Rakahanga and Manihiki, historical records indicate that it maintained close contact with Rarotonga (Southern Cook) for nearly seven decades (1864-1929) before its calendric list was collected \cite{hiroa1932tgv}. Thus, the possibility that the Tongarevan calendar reflects Rarotongan interference must be considered. Continued contact with Rarotongan-speaking communities may also explain why a recent source registers the first lunar night in Tongareva as Tiroe (instead of older Tireo), a form otherwise only attested in the Atiu atoll (Southern Cook; see list 27 in S1A Appendix). 

\subsubsection*{Computational comparison of pairs of names of nights}

To quantify similarity/difference between forms of night names in any pair of calendric lists, we applied an algorithmic method that yielded numerical distances (see below). First, we introduced an adapted version of the Edit Distance metric to compare single forms of names. To find the probable cognate names of nights across pairs of lists, we produced a matching between nights with the lowest cumulative distance (i.e. the sum of all distances between matched nights). This yielded a first measure termed `lexical similarity' ($\mathcal{L}ex$). Second, the algorithm analyzed how lexically cognate nights differ in their relative positions in any pair of lists, to compute the lists' structural divergence ($\mathcal{S}truct$). The algorithm considers the cyclical pattern of lunar calendars: for example, if two nights are found in the 29th and 30th positions in one list, but their counterparts in another list occur in the 30th and 1st positions, respectively, this means that their relative positions are the same. Lastly, we combined the measures $\mathcal{L}ex$ and $\mathcal{S}truct$ to obtain the final distance between any pair of lists. In the next step, we applied the Neighbor-Joining algorithm \cite{saitou&nei1987neighbor} to the distance matrix containing all scores to generate a phylogenetic dendrogram. The resulting tree was rooted using the MAD (Minimal Ancestor Deviation) algorithm \cite{Tria2017}, which has been found to outperform other rooting methods based on experiments with biological phylogenies (Fig. \ref{fig:trees}).

We represented `forms' (i.e. names or parts of names of calendric nights, as collected in S1B Appendix and S2 Dataset) as $n=\langle n_1,...,n_w\rangle$. To accurately identify cognate names of nights, even in cases where linguistic phenomena changed the syntactical order of forms within a name (e.g. Tuamotuan \textit{Hania fakaoti} and \textit{Fakaoti-Hania}), we processed every possible permutation: $P^n=\{P_1^n,...P_{w!}^n\}$. As the name of a night is comprised of only 1 to 4 forms, combinatorial explosion was not an issue. We compared pairs of names using a variation of the well-known Edit Distance, as defined in Figure \ref{fig:edequation}.

\begin{figure*}[ht]
\[
\hspace*{-0.25cm}ED_{nx,ny}(i,j)= 
\begin{cases}
\max(i,j),& \hspace{-2em}min(i,j)=0\\
\min
\begin{cases}
ED_{nx,ny}(i-1,j)+wD\\
ED_{nx,ny}(i,j-1)+wI\\
ED_{nx,ny}(i-1,j-1)+WS(nx_i,ny_j)\cdot wS\\
\end{cases} & \hspace{-2em}min(i,j)\neq0 \\
\end{cases}
\]
\[
WS(nx_i,ny_j)=
\begin{cases}
0 & nx_i=ny_j \\
\multirow{2}{*}{$1-\frac{\sum_{s,t} CS(nx_i^s,ny_j^t)}{max(S,T)}$} & nx_i=nx_i^1\#...\#nx_i^S\\
& ny_j=ny_j^1\#...\#ny_j^T\\
1 & \text{otherwise} \\
\end{cases}
\]
\[
CS(A,B)=
\begin{cases}
0 & A=B \\
\multirow{2}{*}{$0.5$} & A=a_1/a_2, B=b_1/b_2\\
& a_1=b_1\ \&\ a_2\neq b_2 \\
1 & \text{otherwise}\\
\end{cases}
\]
\caption{Definition of the Edit Distance: $wD$, $wI$, and $wS$ represent the weight penalties for sign deletion, insertion, and substitution, respectively. In this study, $wD=wI=1$ and $wS=2$.}
\label{fig:edequation}
\end{figure*}

The function was extended to compute similarity between forms (notating them as described above). The general Edit Distance, including our variation, does not consider word length and is not ideal for comparing the distance between sets of words. Thus, various studies have introduced a form of normalization for Edit Distance values. We relied on the valuable properties \cite{FismanETAL2022} of the \textit{Generalised Edit Distance} (as proposed by \cite{LiLiu2007}), which make it simple and quick to calculate: it is a metric, its upper bound is 1, and it does not escalate repetitions. Thus, we normalized $ED_{nx,ny}$ as
\[
\overline{ED}_{nx,ny}=\frac{2\cdot ED_{nx,ny}}{|nx|+|ny|+ED_{nx,ny}}
\]
where $|\cdot |$ represents the number of compounded forms, separated by the symbol `\_', and forming the name of a night (e.g. $|$faka1\#\textglotstop oti\#\textipa{N}a1\_mase\textglotstop a/samia$|=2$). Given also the need to address trivial grammatical phenomena (including word-order changes), we defined the distance between two nights as
\[
ND_{nx,ny} = min_{P_i^{nx},P_j^{ny}}\ \  \overline{ED}_{P_i^{nx},P_j^{ny}}
\]
searching for the pair of permutations of the two names of nights that have the minimal distance.

\subsubsection*{Lexical measure}
Relations of cognacy between names of nights from two calendric lists may entail 1-to-1 mappings between cognate nights. To address this possibility and facilitate accurate evaluation, for calendar matching we relied on the standard \textit{Linear Sum Assignment} (LSA) problem, which can be solved by the \textit{Hungarian algorithm}. We defined $C_X = \langle nx_1,...,nx_K\rangle$ and $C_Y = \langle ny_1,...,ny_M\rangle$, that is, any two calendric lists to be compared, as two tuples of nights. We also assumed that $K\leq M$ , without any loss of generality, enabling the comparison of lists with different numbers of items. We can define a matching solution $\boldsymbol{\Xi}_{C_X,C_Y}=\langle \xi_{i,j}\rangle$ as a matrix over the variables $\xi_{i,j}\in\{0,1\}$ representing the calendar alignment obtained by the LSA algorithm, with $\xi_{i,j}=1$ if and only if $nx_i$ is assigned to $ny_j$ and 0 otherwise. The LSA problem to be solved can then be expressed as
\[
\mathcal{L}ex_{C_X,C_Y} = \min\mspace{-2mu}\sum_{i=1}^{K} \sum_{j=1}^{M} \xi_{i,j}\cdot ND_{nx_i,ny_j}
\]
\begin{align*}
\textrm{s.t.}\ \ \  & \sum_{i=1}^{K} \xi_{i,j}\leq 1,\ \ \  j=1,2,...,M\\
     & \sum_{j=1}^{M} \xi_{i,j}=1,\ \ \  i=1,2,...,K
\end{align*}

Note that: if $K\leq M$, then $M-K$ nights are not matched.

\subsubsection*{Structural measure}

The Lee Distance for two general permutations of integers $p_1$ and $p_2$ of length $n$ is:
\[
\delta(p1,p2)=\sum_{e\in p_1} min(|i-j|,K-|i-j|), \ \textrm{where}\ p_1(i)=p_2(j)=e
\]
Considering that a LSA matching can be expressed as two permutations of positions of nights in $C_X$ and $C_Y$, it can be adapted to our problem as
\[
\delta(\boldsymbol{\Xi}_{C_X,C_Y})=\sum_{\xi_{i,j}=1} min(|i-j|,K-|i-j|).
\]

and then (following \cite{BariffiETAL2022}) normalized as
\[
\overline{\delta(\boldsymbol{\Xi}_{C_X,C_Y})}=\frac{\delta(\boldsymbol{\Xi}_{C_X,C_Y})}{(K*\lfloor M/2 \rfloor)}.
\]

In some instances, two lists are identical except for the presence of an extra night (e.g. RPN1 and RPN3). In these cases, $\delta(\boldsymbol{\Xi}_{C_X,C_Y})$ is not a faithful measure as it counts a single shift multiple times. Thus, our method fixed the first list and shifted the nights of the second list in search for the best rotational matching: 
\[
\delta(\boldsymbol{\Xi}_{C_X,C_Y})_k=\sum_{\xi_{i,j}=1} min(|i-(j+k)|,K-|i-(j+k)|).
\]
Moreover, typically there was not a single best matching solution for any pair of lists, so we had to explore a set of equivalent solutions having the same $\mathcal{L}ex$ value but different matching $\boldsymbol{\Xi}_{C_X,C_Y}$, and forming a set of possible solutions $\{\boldsymbol{\Xi}_{C_X,C_Y}^z\}$.

The final structural distance $\mathcal{S}truct$ can thus be defined as
\[
\mathcal{S}truct_{C_X,C_Y} = min_z \left[min_{k}\ \overline{\delta(\boldsymbol{\Xi}_{C_X,C_Y}^z)_k}\right].
\]

\subsubsection*{Calendric distance}

The final calendric distance combined the lexical ($\mathcal{L}ex$) and the structural ($\mathcal{S}truct$) measures. We did not simply sum their contributions. If two lists are very different, then their $\mathcal{L}ex$ score would be high—close to $1$—and most of the distances between individual pairs of nights would also be high. This means that the LSA matching will, correctly, produce a high number of equivalent but poorly matching solutions, with nights matched almost randomly. If we searched for the best matching through double minimization in the definition of $\mathcal{S}truct$, we would certainly find a suitable solution with a very low $\mathcal{S}truct$ score. This is evidently incongruous: the meaningfulness of the measure $\mathcal{S}truct$ had to be modulated in some way to avoid such degenerate behaviour. Given that the contribution of the $\mathcal{S}truct$ distance must be maximal for $\mathcal{L}ex$ near 0 and minimal for $\mathcal{L}ex$ near 1, for deriving the final distance between two calendars  $C_X$ and $C_Y$ we combined the two measures as
\[
\mathcal{D}ist_{C_X,C_Y} = \mathcal{L}ex_{C_X,C_Y} + (1-\mathcal{L}ex_{C_X,C_Y}) \cdot \mathcal{S}truct_{C_X,C_Y}
\]

By combining the metrics described above to derive $\mathcal{D}ist_{C_X,C_Y}$, we were then able to compare every calendrical list and produce a distance matrix suitable for the application of the Neighbor-Joining clustering method and the Minimal Ancestor Deviation rooting algorithm (as described above).


\subsubsection*{Exclusion of other calendric types}

We did not explore the divergence between East-Polynesian `naming' lists and West-Polynesian `numbering' lists, although some of the latter include a few related forms (e.g., there are cognates of *\textit{tuu}, *\textit{matofi}, *\textit{faka\textglotstop oti}, *\textit{mate}, etc.; see S1C Appendix). Three reasons hinder the application of our comparative method to a set containing both types. First, while we have complete records of lists for several locations that use the `numbering' type (Pukapuka, Rennell and Bellona, Samoa, Takuu, Tokelau, Tonga, and Vaitupu), in other instances we located only partial data (Nanumea, Niue, and Sikaiana). Second, in some cases nights were described with long sentences (e.g., the 28th night in the Pukapukan list is \textit{Koa wakatau wenake ma te ulu o te ata matua} ``[The moon] rises at the same time as the darkness before dawn'') making it difficult to isolate diagnostic and comparable lexical forms. Third, and most importantly, number words are pervasive in most West-Polynesian lists, with some sequences of nights numbered from `one' to as many as `ten'. This creates a risk for our method, as it might falsely indicate greater similarity with the East-Polynesian calendars that have longer numbered series (such as those from Rapa Nui), which may be an independent innovation.

We also did not apply our method to the year calendar used in both West and East Polynesia, which had months also with specific names \cite{best1922maori, hiroa1938mgv, kirch2001hawaiki}. This calendar has been the focus of comparative studies and is found in similar forms across several archipelagos, but some locations lack comprehensive documentation or show disruptions. Notably, the Marquesan lists suffer from transcription issues, and the surviving Rapa Nui month names have minimal to no similarities with other Polynesian calendars. Moreover, establishing cognacy and patterns of divergence is challenging due to the practice of renaming months after the stars marking their beginning in certain islands.

\section*{Results and Discussion}

\begin{figure*}
     \centering
     \begin{subfigure}{\textbf{A}}
         \centering
         \fbox{\includegraphics[width=0.85\textwidth]{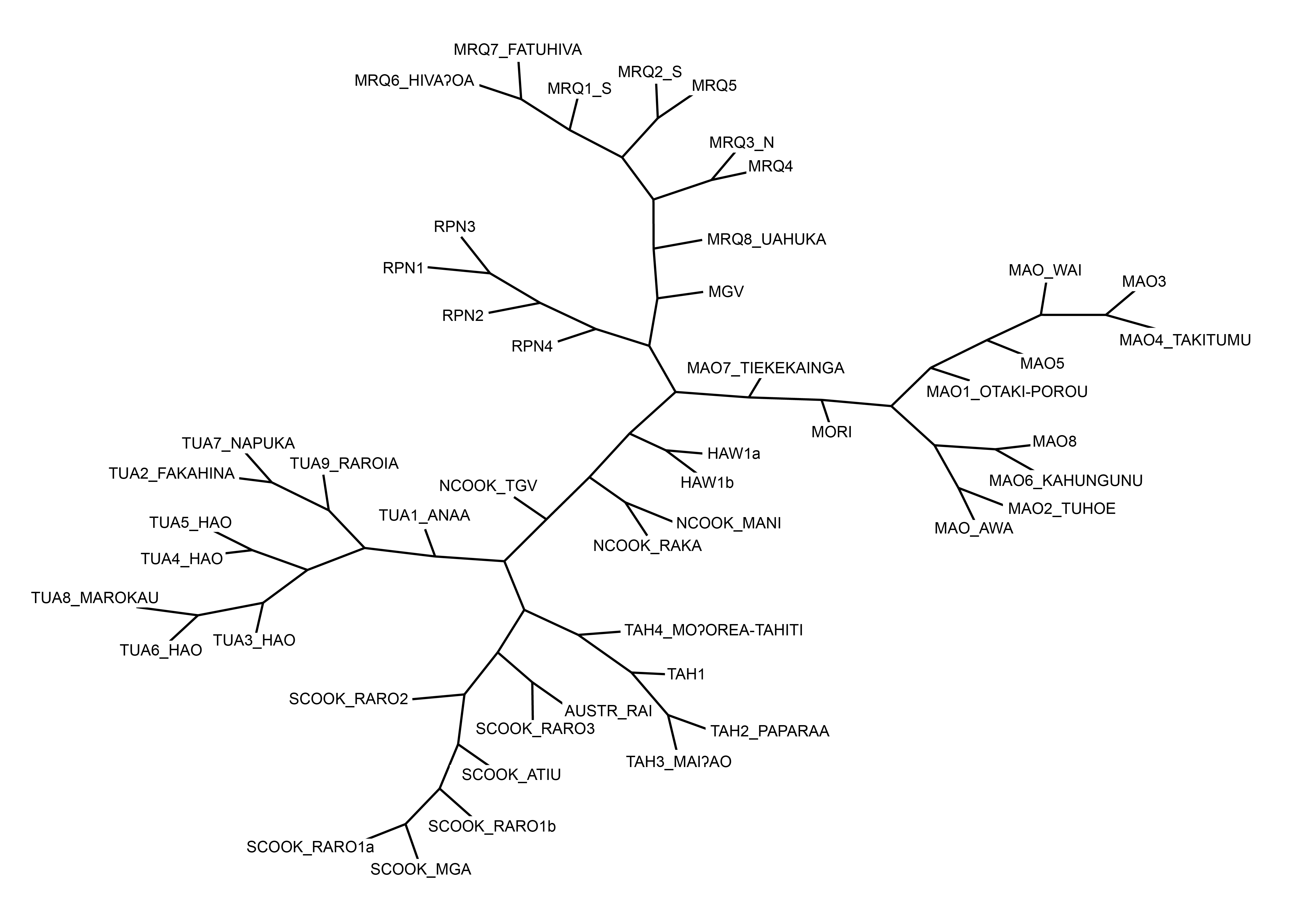}}
         \label{fig:2a}
         \newline
     \end{subfigure}
     \hfill
     \begin{subfigure}{\textbf{B}}
         \centering
         \fbox{\includegraphics[width=0.85\textwidth]{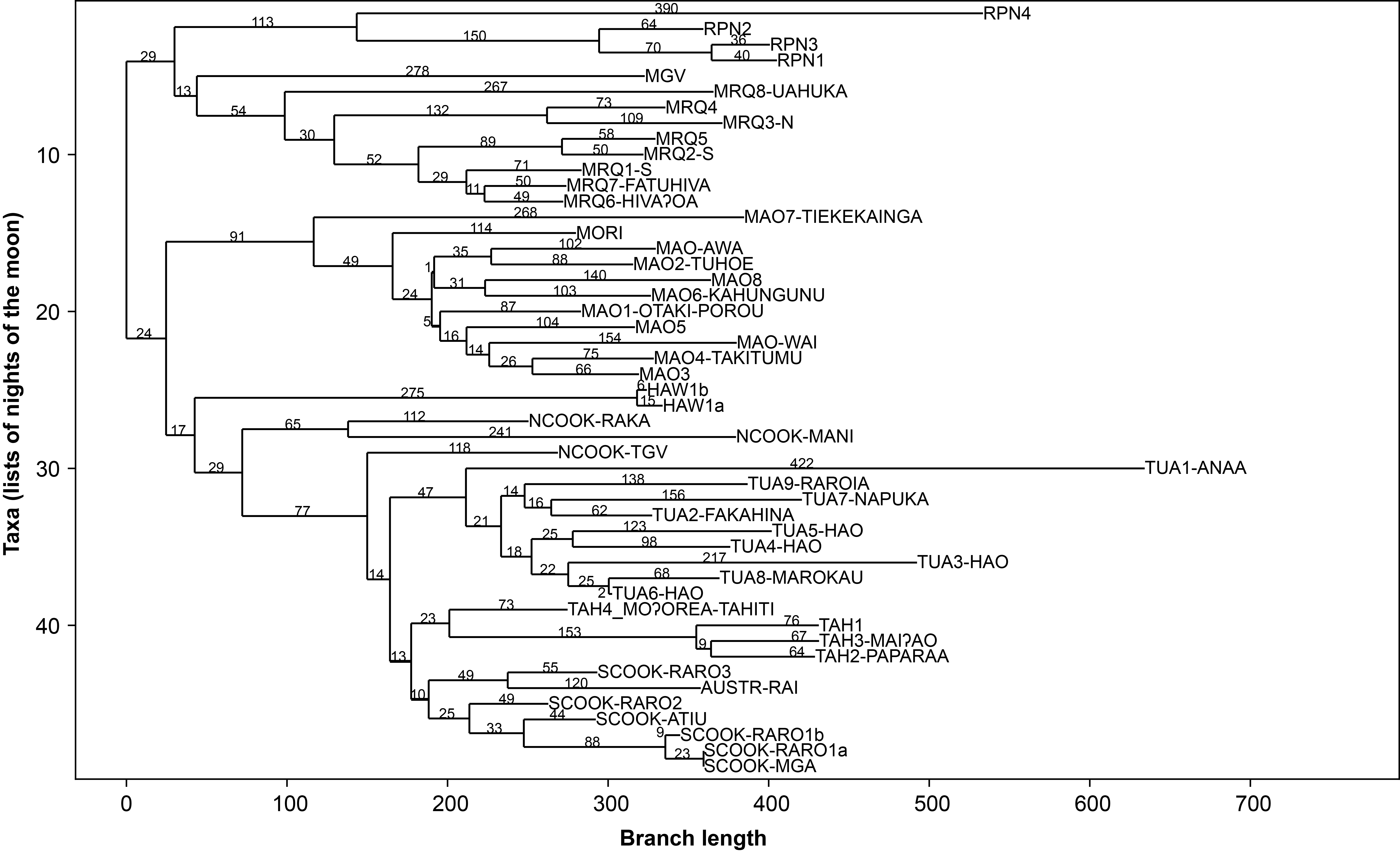}}
        \label{fig:2b}
        \newline
     \end{subfigure}
     \caption{Phylogeny of 49 lists of ’the nights of the moon’ from East Polynesia: unrooted (\textbf{A}) and
rooted using the MAD algorithm (\textbf{B}). Key to the macro-provenances (islands or archipelagos)
of the calendric lists: AUSTR: Austral Islands; HAW: Hawai'i; MAO: Aotearoa/New
Zealand; MGV: Mangareva (in the Gambier archipelago) MRQ: Marquesas Islands; MORI:
Moriori/Chatham Islands; NCOOK: Northern Cook Islands; SCOOK: Southern Cook Islands;
RPN: Rapa Nui/Easter Island; TAH: Tahiti and neighboring islands (Society Islands, French
Polynesia); TUA: Tuamotu Islands. See S1A Appendix for the specific source of each list.}
        \label{fig:trees}
\end{figure*}

Our tree (Fig. \ref{fig:trees}B) accurately reflects the similarity between lunar night lists from geographically close areas, validating the results. The sets from New Zealand, the Tuamotuan archipelago, Tahiti (and nearby islands), the Marquesas, and the southern Cook were all grouped together in specific branches or sub-branches. Moreover, the four lists from the Tuamotuan atoll of Hao not only appear as closely related `taxa', but also share a common node (ancestor) with the list from the neighboring atoll of Marokau. The Maori lists attributed to the Awa and Tuhoe tribal groups, both from the northeastern part of North Island (New Zealand), also branch out from the same node. The lists from the southern and northern Marquesas, respectively, are located in separate sub-branches of the tree. Our method also succeeded in recognizing the correct phylogenesis of one `corrupted' list from Rapa Nui (the one we labeled RPN4) and two lists reconstructed from relatively late dictionary entries (TUA1, RARO3; see S1A Appendix for details). However, as also expected, the tree places these very divergent lists (RPN4, TUA1) at considerable distances to other `taxa' with common nodes (i.e. `ancestor' lists). 

Furthermore, our phylogeny aligns with the real distribution of shared vs non-shared night names. The tree has two main branches or groups of calendric lists (Branch 1 and Branch 2), consistent with an ancient two-way split of the proto-lunar calendar. Branch 1 includes the lists from the Marquesas, Mangareva, and Rapa Nui, while Branch 2 encompasses the calendars from New Zealand, Hawai'i, the Cook and Austral islands, Tahiti, and the Tuamotu. We observe two key distinctions between the lists in these larger branches. First, those in Branch 1 feature a single \textit{*Raakau} night, either before or after a night called \textit{*Matofi}, whereas those in Branch 2 have two or more \textit{*Raakau} nights forming a series and found alongside \textit{*Matofi} only sometimes. Second, none of the lists in Branch 1 contains a night whose name is a reflex of *\textit{Tamatea}. Conversely, all sub-ramifications of Branch 2 include a night or series of nights named \textit{*Tamatea}, except for those from Hawai'i (although it is possible that \textit{*Tamatea} originally existed in Hawai'i and was later substituted and dropped). When we consider further subdivisions of the tree, we find that in general they agree well with shared or non-shared features among calendars. Another advantageous property that emerged is that the ancestor of the Hawaiian lists is placed the closest to the ancestor of the Rakahanga and Manihiki ones, which agrees with their geographical positions (Fig. \ref{fig:map}) and a scenario in which Hawai'i was settled by migrants from islands located to the southwest of the archipelago.

Only the sub-branch with the Tuamotuan, Tongarevan, southern Cook and Tahitian lists poses issues. In these lists night series are ordered using \textit{*roto} `middle', probably a shared innovation. Yet, while most follow the pattern \textit{*tahi, *roto, }*(\textit{faka})\textit{\textglotstop oti} `one, inside, final', three of the Tahitian lists use \textit{*mu\textglotstop a, *roto, *muri} `before, inside, after'. In addition, the sets from Tongareva and Tahiti have a night called \textit{Ari}. This is shared with other sub-branches of Branch 2 as well as Branch 1, and must therefore be a retention (`archaism'). Conversely, the Tuamotu and southern Cook sets have \textit{Vari} instead of \textit{Ari}, likely an innovation (as noted in the Introduction). These features differentiate the Tahitian lists from those of the Tuamotu and southern Cook, but the tree does not imply the separate evolution of the Tahitian sets. Instead, it produces a common node (ancestor) for the Tahitian and southern Cook lists, and separates the Tuamotuan calendars. This result is not unexpected: as noted above, Tahiti exerted influence over neighboring archipelagos from the late 18th century onward, within a central area of East Polynesia that includes some of the least isolated islands. As a result, horizontal influence among calendars in this region is more likely. This may explain why the sub-branches do not show the Tahitian lists as more distinct from the others. Similarly, the fact that the Tongarevan list is not more closely aligned with the Rahakanga and Manihiki lists—despite their geographical proximity within the Northern Cook Islands—and instead clusters with the Tahitian, Tuamotuan, and Southern Cook sub-branch, is likely due to historically documented contact with populations from Rarotonga (Southern Cook) in the 19th century. 

The sequence of settlement in East Polynesia is likely reflected—at least to some degree—in the historical development of both East Polynesian lunar calendars and languages, as both are symbolic systems transmitted and changed within communities. Although long-distance sea travel in the pre-industrial Pacific facilitated the settlement of Polynesia, it is important to remember the vast distances and significant isolation of many islands and archipelagos in the region, which greatly limited contact and borrowing. In this context, the current phylogeny of the lunar calendars offers implications worth considering.

As regards the historical development of East Polynesian languages, the most cited classification proposes a Proto-Central-Eastern Polynesian language, which first divided into Rapanui and Proto-East Polynesian. Proto-East Polynesian is subsequently divided into two major sub-groups: `Tahitic' (including Tahitian, New Zealand Maori, Tuamotuan, etc.) and `Marquesic' (comprising Hawaiian, Mangarevan, and Marquesan) \cite{green1966}. More recently, two alternative classifications have been proposed. One, put forward by Walworth, suggests that Proto-East Polynesian first split into an isolated Rapa Nui branch and a Central Eastern Polynesian (CEP) group comprising all other languages. According to this proposal, the CEP languages did not divide into Marquesic and Tahitic groups as traditionally postulated, but rather developed through ongoing waves of contact driven by alleged high inter-island mobility, spreading linguistic features even to distant regions such as Aotearoa/New Zealand and Hawai‘i \cite{walworth2014EPrev}. The other alternative view, proposed by Wilson, suggests that Proto-East Polynesian split into two groups: a 'Distal' group, including Marquesan, Mangarevan, and Rapa Nui, and a 'Proximal' group, comprising the remaining languages, including Hawaiian \cite{wilson2021EPsub}.

\begin{figure}[ht]
    \centering
    \fbox{\includegraphics[width=0.75\linewidth]{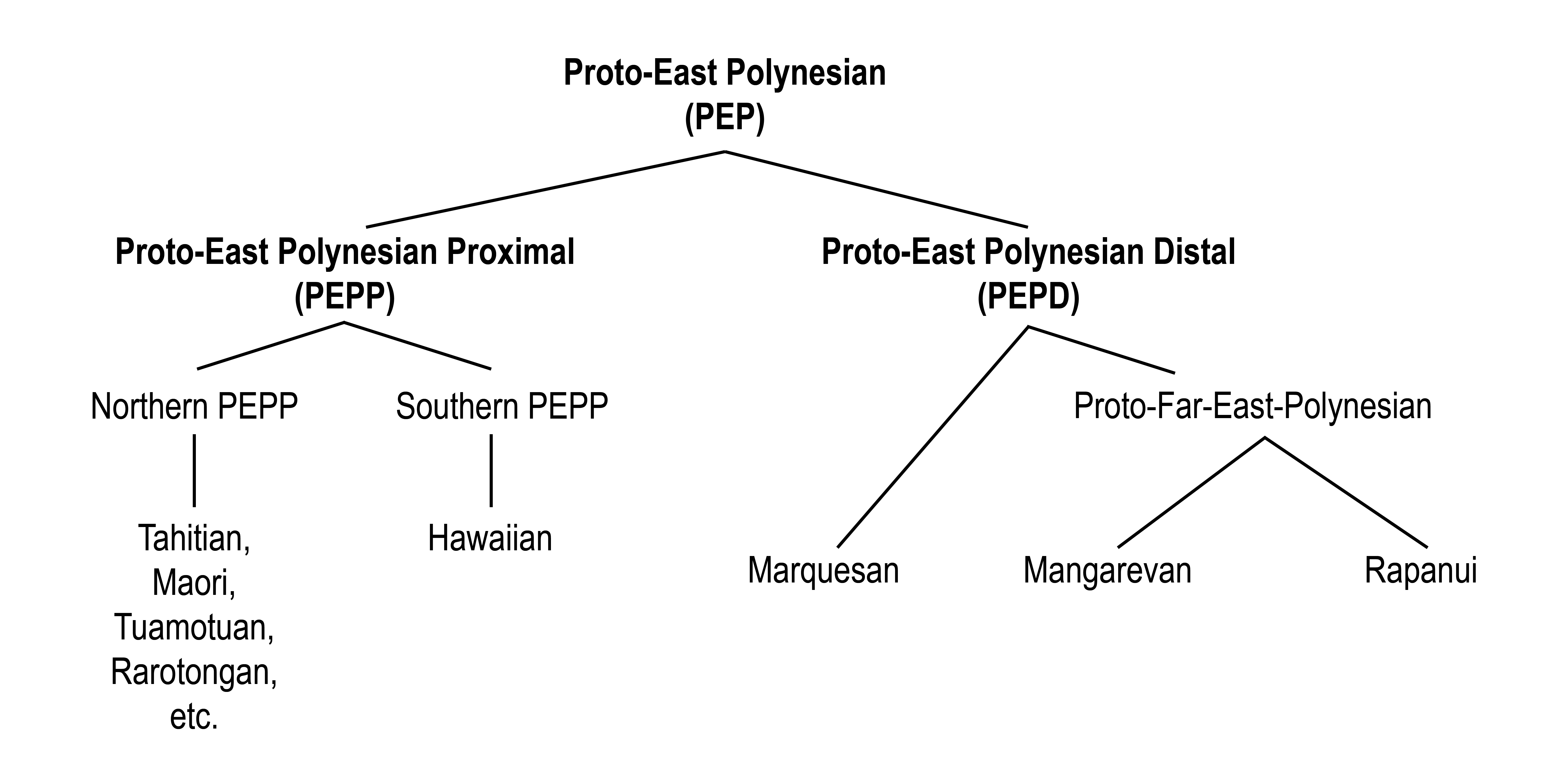}}
    \caption{Wilson's alternative subgrouping of East-Polynesian languages into Proto-East Polynesian ‘Proximal’ and ‘Distal’ divisions (adapted from  \cite{wilson2021EPsub}).}
    \label{fig:alt-subgrouping}
\end{figure}

As it happens, our tree aligns with the division of East-Polynesian languages in `Distal' and `Proximal' sub-groups. According to this classification, `Distal' languages include Marquesan (all dialects), Mangarevan, and Rapanui, while `Proximal' languages comprise Hawaiian, Mangaian (southern Cook), Manihiki-Rakahanga, Moriori, New Zealand Maori, Tongarevan/Penrhyn, (Old) Rapan, Rarotongan (southern Cook), Tahitian, and Tuamotuan \cite{wilson2021EPsub} (Fig. \ref{fig:alt-subgrouping}). 
The archipelagos where these two linguistic subgroups were spoken closely match those where the lists in the two main branches of our calendric phylogeny (those nearest the root of the tree) were used. Accordingly, our findings can be interpreted as evidence that the earliest major divergences in the East Polynesian proto-language and proto-lunar calendar occurred in parallel, likely reflecting an ancient two-way movement of populations. In this scenario, speakers of ‘Proto-East Polynesian Distal’ used a lunar calendar ancestral to the Rapa Nui, Mangarevan, and Marquesan sets (the node of Branch 1), and their subsequent dispersal to these islands resulted in the parallel divergence of both their language and their lunar calendar. Likewise, we might infer that speakers of `Proto-East Polynesian Proximal' used the ancestor to the remaining lists (node of Branch 2).

However, after this point our results do not necessarily align with Wilson's linguistic phylogeny anymore. One tenet of his linguistic subgrouping is that `Proximal' languages initially divided into `Southern' and `Northern' dialects, with the latter being ancestral to Hawaiian and the former giving rise to all other `Proximal' tongues \cite{wilson2021EPsub}. Conversely, our dendrogram indicates a greater divergence of the Maori (New Zealand) and related Moriori (Chatham) calendric lists with regard to the rest of areas where `Proximal' tongues were spoken. Thus, in the future, linguistic data might be reexamined to reconsider the sub-subgrouping of the Maori and Hawaiian languages. 

\section*{Conclusions}

The phylogeny of East Polynesian lists of 'nights of the moon' presented here aligns closely with one proposed classification of East Polynesian languages, strongly suggesting parallel historical developments in both symbolic systems. We conclude that this correspondence is unlikely to be due to chance and reflects—at least to some extent—the paths of human settlement in East Polynesia, with a probable two-way population split: one leading to the settlers of the Marquesas, Mangareva, and Rapa Nui, and the other to the settlers of the remaining East Polynesian-speaking archipelagos. This challenges previous models which propose that Rapanui was the first linguistic community to diverge from the rest. Nevertheless, we emphasize the independent nature of the evidence provided by the computational analysis of lexical and structural similarities across nearly 50 calendric lists, examined systematically here for the first time. Its results offer new insights into the settlement history of East Polynesia and should be considered alongside archaeological, genetic, and linguistic data. 

\bibliography{biblio}

@inproceedings{FismanETAL2022,
  author    = {Dana Fisman and
               Joshua Grogin and
               Oded Margalit and
               Gera Weiss},
  editor    = {Hideo Bannai and
               Jan Holub},
  title     = {{The Normalized Edit Distance with Uniform Operation Costs Is a Metric}},
  booktitle = {33rd Annual Symposium on Combinatorial Pattern Matching, {CPM} 2022,
               June 27-29, 2022, Prague, Czech Republic},
  series    = {LIPIcs},
  volume    = {223},
  pages     = {17:1--17:17},
  publisher = {Schloss Dagstuhl - Leibniz-Zentrum f{\"{u}}r Informatik},
  year      = {2022},
}

@article{LiLiu2007,
  author={Li, Yujian and Liu, Bo},
  journal={IEEE Transactions on Pattern Analysis and Machine Intelligence}, 
  title={{A Normalized Levenshtein Distance Metric}}, 
  year={2007},
  volume={29},
  number={6},
  pages={1091-1095},
}

@book{best1922maori,
  title={{The Maori Division of Time}},
  author={Best, E.},
  lccn={24008294},
  series={Dominion Museum monograph},
  year={1922},
  publisher={Dominion Museum}
}

@incollection{biggs1971,
    author = {Biggs, Bruce},
    title = {{The Languages of Polynesia}},
    booktitle = {{Linguistics in Oceania}},
    publisher = {De Gruyter},
    year = {1971}
}

@book{craig1989dictionary,
  title={{Dictionary of Polynesian Mythology}},
  author={Craig, R.D.},
  isbn={9780313258909},
  lccn={89007479},
  year={1989},
  publisher={Bloomsbury Academic}
}

@book{denoon2004cambridge,
  title={{The Cambridge History of the Pacific Islanders}},
  author={Denoon, D.},
  isbn={9780521003544},
  lccn={96052784},
  series={Cambridge Histories Online},
  year={2004},
  publisher={Cambridge University Press}
}

@book{hiroa1932maniraka,
  title={{Ethnology of Manihiki and Rakahanga}},
  author={Hiroa, T.R.},
  series={Bernice P. Bishop Museum Bulletin},
  year={1932},
  publisher={Bernice P. Bishop Museum}
}

@book{hiroa1932tgv,
  title={{Ethnology of Tongareva}},
  author={Hiroa, T.R.},
  lccn={33013688},
  series={Bernice P. Bishop Museum bulletin},
  year={1932},
  publisher={Bernice P. Bishop Museum}
}

@book{hiroa1938mgv,
  title={{Ethnology of Mangareva}},
  author={Hiroa, T.R.},
  lccn={39013600},
  series={Bernice P. Bishop Museum Bulletin},
  year={1938},
  publisher={Bernice P. Bishop Museum}
}

@book{burrows1938western,
  title={{Western Polynesia, a Study in Cultural Differentiation}},
  author={Burrows, E.G.},
  lccn={39021391},
  series={Ethnological studies},
  year={1938},
  publisher={Kaudern}
}

@article{green1966,
 ISSN = {00324000},
 author = {Roger Green},
 journal = {The Journal of the Polynesian Society},
 number = {1},
 pages = {6--38},
 publisher = {Polynesian Society},
 title = {{Linguistic subgrouping within Polynesia: The implications for prehistoric settlement}},
 volume = {75},
 year = {1966}
}

@book{green1985,
  address={Auckland},
  author={Green, Roger C.},
  number={68},
  publisher={University of Auckland},
  series={Working Papers in Anthropology, Archaeology, Linguistics, Maori Studies},
  booktitle={{Subgrouping of the Rapanui language of Easter Island in Polynesian and its implications for East Polynesian prehistory}},
  title={{Subgrouping of the Rapanui language of Easter Island in Polynesian and its implications for East Polynesian prehistory}},
  year={1985}
}

@article{horley2011,
  title={{Lunar calendar in rongorongo texts and rock art of
Easter Island}},
  author={Horley, Paul},
  journal={Journal de la Soci{\'e}t{\'e} des Oc{\'e}anistes},
  volume={132},
  pages={17-38},
  year={2011}
}

@article{ioannidis2021,
  author = {Alexander G. Ioannidis and Javier Blanco-Portillo and Karla Sandoval and Erika Hagelberg and Carmina Barberena-Jonas and Adrian V. S. Hill and Juan Esteban Rodríguez-Rodríguez and Keolu Fox and Kathryn Robson and Sonia Haoa-Cardinali and Consuelo D. Quinto-Cortés and Juan Francisco Miquel-Poblete and Kathryn Auckland and Tom Parks and Abdul Salam M. Sofro and María C. Ávila-Arcos and Alexandra Sockell and Julian R. Homburger and Celeste Eng and Scott Huntsman and Esteban G. Burchard and Christopher R. Gignoux and Ricardo A. Verdugo and Mauricio Moraga and Carlos D. Bustamante and Alexander J. Mentzer and Andrés Moreno-Estrada},
  title = {{Paths and timings of the peopling of Polynesia inferred from genomic networks}},
  journal = {Nature},
  volume = {597},
  number = {7877},
  pages = {522--526},
  year = {2021},
  doi = {10.1038/s41586-021-03902-8},
  url = {https://doi.org/10.1038/s41586-021-03902-8}
}

@book{kirch2001hawaiki,
  title={{Hawaiki, Ancestral Polynesia: An Essay in Historical Anthropology}},
  author={Kirch, P.V. and Green, R.C.},
  isbn={9780521788793},
  lccn={00031175},
  year={2001},
  publisher={Cambridge University Press}
}

@article{pawley1966,
 author = {Pawley, Andrew},
 journal = {The Journal of the Polynesian Society},
 number = {1},
 pages = {39-64},
 publisher = {Polynesian Society},
 title = {{Polynesian languages: a subgrouping based on shared innovations in morphology}},
 urldate = {2024-03-24},
 volume = {75},
 year = {1966}
}

@book{marck2000topics,
  title={{Topics in Polynesian Language and Culture History}},
  author={Marck, J.C.},
  isbn={9780858834682},
  lccn={00227226},
  series={Pacific linguistics},
  year={2000},
  publisher={Pacific Linguistics, Research School of Pacific and Asian Studies, The Australian National University}
}

@book{metraux1940,
  title={{Ethnology of Easter Island}},
  author={M{\'e}traux, A.},
  lccn={40029585},
  series={Bernice P. Bishop Museum},
  year={1940},
  publisher={Bernice P. Bishop Museum}
}

@misc{POLLEX,
    title = {{POLLEX}},
    url = {https://pollex.eva.mpg.de/},
    author = {S.J. Greenhill  and R. Clark},
    year = {2011},
    note = {Accessed on July 28th, 2024},
    urldate = {2024-03-24},
}

@article{saitou&nei1987neighbor,
  title={{The Neighbor-joining Method: A New Method for
Reconstructing Phylogenetic Trees}},
  author={Saitou, Naruya and Nei, Masatoshi},
  journal={Molecular Biology and Evolution},
  volume={4},
  number={4},
  pages={406--425},
  year={1987},
  month={Jul},
  doi={10.1093/oxfordjournals.molbev.a040454},
  PMID={3447015}
}

@article{stimson1928,
 ISSN = {00324000},
 URL = {http://www.jstor.org/stable/20702217},
 author = {J. Frank Stimson},
 journal = {The Journal of the Polynesian Society},
 number = {3(147)},
 pages = {326-337},
 publisher = {Polynesian Society},
 title = {{Tahitian names for the nights of the moon}},
 volume = {37},
 year = {1928}
}

@article{stimson1930hamzah,
  author    = {Stimson, J. F.},
  title     = {A Discussion of the Hamzah and Some Allied Aspects of Polynesian Phonetics},
  journal   = {Journal of the Polynesian Society},
  volume    = {39},
  number    = {3},
  pages     = {263--283},
  year      = {1930}
}

@article{Tria2017,
author={Tria, Fernando Domingues K{\"u}mmel
and Landan, Giddy
and Dagan, Tal},
title={{Phylogenetic rooting using minimal ancestor deviation}},
journal={Nature Ecology {\&} Evolution},
year={2017},
month={Jun},
day={19},
volume={1},
number={7},
pages={0193},
issn={2397-334X},
doi={10.1038/s41559-017-0193},
url={https://doi.org/10.1038/s41559-017-0193}
}

@article{valerio&al2022,
  issn={1720-9331},
  title={{The Rongorongo Tablet C: new technologies and conventional approaches to an undeciphered text}},
  number={2},
  author={Miguel Val{\'e}rio  and Lorenzo Lastilla  and Roberta Ravanelli},
  journal={Lingue e Linguaggio},
  volume={21},
  pages={333–367},
  year={2022}
}

@article{
wilmshurt&al2011dating,
author = {Janet M. Wilmshurst  and Terry L. Hunt  and Carl P. Lipo  and Atholl J. Anderson },
title = {{High-precision radiocarbon dating shows recent and rapid initial human colonization of East Polynesia}},
journal = {Proceedings of the National Academy of Sciences},
volume = {108},
number = {5},
pages = {1815-1820},
year = {2011},
doi = {10.1073/pnas.1015876108},
URL = {https://www.pnas.org/doi/abs/10.1073/pnas.1015876108},
eprint = {https://www.pnas.org/doi/pdf/10.1073/pnas.1015876108},
}

@article{walworth2014EPrev,
 ISSN = {00298115, 15279421},
 URL = {http://www.jstor.org/stable/43286530},
 author = {Walworth, Mary},
 journal = {Oceanic Linguistics},
 number = {2},
 pages = {256-272},
 publisher = {University of Hawai'i Press},
 title = {{Eastern Polynesian: The Linguistic Evidence Revisited}},
 urldate = {2024-03-24},
 volume = {53},
 year = {2014}
}

@article{wilson2021EPsub,
 URL = {https://doi.org/10.1353/ol.2021.0001},
 author = {Wilson, William H.},
 journal = {Oceanic Linguistics},
 number = {1},
 pages = {36-71},
 publisher = {},
 title = {{East Polynesian Subgrouping and Homeland Implications Within the Northern Outlier–East Polynesian Hypothesis}},
 volume = {60},
 year = {2021}
}

@inproceedings{BariffiETAL2022,
	year = {2022},
	booktitle = {{International Zurich Seminar on Information and Communication (IZS 2022). Proceedings}},
	editor = {Lapidoth, Amos and Moser, Stefan M.},
	author = {Bariffi, Jessica and Bartz, Hannes and Liva, Gianluigi and Rosenthal, Joachim},
	address = {Zurich},
	publisher = {ETH Zurich},
	title = {{On the Properties of Error Patterns in the Constant Lee Weight Channel}},
	pages = {44 - 48},
}

@incollection{valerio2024rongorongo,
  author    = {Valério, Miguel},
  title     = {{The Rongorongo ‘Lunar Calendar’ of Rapa Nui (Easter Island) and the Type of Script}},
  booktitle = {Writing from Invention to Decipherment},
  editor    = {Ferrara, S. and Montecchi, B. and Valério, M.},
  year      = {2024},
  publisher = {Oxford University Press},
  address   = {Oxford},
  pages     = {189--224}
}

\end{document}